\documentclass[12pt,preprint]{aastex}

\slugcomment{This manuscript is accepted for publication in
\textbf{Ap\&SS}}

\shorttitle{MBES with AD heating}

\shortauthors{Nejad-Asghar}

\begin{document}

\title{Modified Bonnor-Ebert spheres with ambipolar diffusion heating}

\author{Mohsen Nejad-Asghar}

\affil{Department of Physics, University of Mazandaran, Babolsar,
Iran}

\email{nejadasghar@umz.ac.ir}

\begin{abstract}
Magnetic fluctuations through the molecular cloud cores can produce
ambipolar diffusion (AD) heating, which consequently can produce
temperature gradients through the core. The aim of this paper is to
investigate the effects of these produced temperature gradients on
the radius and mass of the non-isothermal modified Bonnor-Ebert
spheres (MBES). Here, we use the parameter $\kappa$ to represent the
magnetic fluctuations through the molecular cloud cores. This
parameter introduces the change of magnetic filed strength in the
length-scale. The results show that increasing of $\kappa$ leads to
an increase of the radius and mass of MBES. The most important
result is existence of the gravitationally stable high-mass
prestellar cores at the low-density molecular medium with great
magnetic fluctuations.
\end{abstract}

\keywords{ISM: clouds -- stars: formation -- hydrodynamics --
instabilities}

\section{Introduction}

The gravitational-unstable and collapsing subset of the molecular
cloud cores are known as prestellar cores (Ward-Thompson et al.
1994), which will convert to an individual stellar system with its
universal initial mass function (Offner et al. 2014). Thus,
investigation of the prestellar cores is essential to understand the
initial stages of star (and stellar system) formation. Many authors
use the usual \textit{isothermal} Bonnor-Ebert sphere (e.g. Stahler
\& Palla~2004, Section~9.1) to approximate the density profiles of
observed prestellar cores (e.g, Marsh et al. 2016). However, in
general, the assumption of \textit{isothermal case} is not
accurately valid because real dense cores are subjected to heating
and cooling processes (Goldsmith \& Langer~1978). For this purpose,
some authors assumed \textit{non-isothermal cases} to approximate
self-consistently the temperature and density profiles through the
thermally balanced hydrostatic spherical prestellar cores (e.g.,
Evans~II et al.~2001, Juvela \& Ysard~2011, Spil\"{a}, Harju \&
Juvela~2015). In these works for non-isothermal modified
Bonnor-Ebert spheres (MBES), the ambipolar diffusion (AD) heating
was not considered, while in the magnetic fluctuating medium, this
heating mechanism is very important (e.g., Li, Myers \& McKee~2012).

Observations of prestellar cores show the effects of magnetic
turbulence through them. For example, Hildebrand et al.~(2009)
developed a method based on a dispersion function about local mean
magnetic fields. Besides providing a measure for the turbulent
dispersion, their method also gives an accurate estimate of the
turbulent to the mean magnetic field strength ratio. Koch, Tang \&
Ho~(2010) applied and extended the method developed in Hildebrand et
al.~(2009) across a range of scales in the same star formation
regions. They found the magnetic field turbulent dispersion, its
turbulent-to-mean field strength ratio, and the large-scale
polarization angle correlation length as a function of the physical
scale at the star formation sites. Girart et al.~(2013) also found
evidence of magnetic field diffusion at the core scales, far beyond
the expected value for AD. They concluded that it may be possible
that the diffusion arises from fast magnetic reconnection in the
presence of turbulence. Similarly, we can also refer to the recent
observational evidences of Pillai et al.~(2015) and Houde et
al.~(2016) for high-mass star-forming regions.

Magnetic turbulence can produce magnetic fluctuations through
prestellar molecular cloud cores. Thus, prestellar cores are a
magnetically fluctuating medium, and the AD heating is important and
must be considered. For this purpose, we consider the physical
situation of a spherically symmetric, gravitationally  bound,
thermally supported prestellar core, which is also heated by
internal magnetic slippage of particles with AD mechanism in the
magnetic fluctuating medium. Section~2 gives the formulation of the
quasi-static thermally equilibrium prestellar cores, and in
section~3 some remarkable conclusions are presented.

\section{Hydrostatic thermally equilibrium cores}

In spherical polar coordinates, the usual hydrodynamic equations for
spherically symmetric thermally dominated cases are
\begin{equation}\label{contin}
    \frac{\partial \rho}{\partial t} + \frac{1}{r^2} \frac{\partial}{\partial
    r} \left( r^2 \rho u \right) = 0,
\end{equation}
\begin{equation}\label{moment}
    \frac{\partial u}{\partial t} + u\frac{\partial u}{\partial r}
    =-\frac{1}{\rho} \frac{\partial p}{\partial r} - \frac{GM}{r^2},
\end{equation}
\begin{equation}\label{mass}
    \frac{\partial M}{\partial t} + u \frac{\partial M}{\partial r}
    = 0,\quad \frac{\partial M}{\partial r} = 4 \pi r^2 \rho,
\end{equation}
\begin{equation}\label{energy}
    \frac{\partial p}{\partial t} + u \frac{\partial p}{\partial r}
    + \gamma p \frac{1}{r^2} \frac{\partial}{\partial r} \left( r^2 u
    \right) = - \left( \gamma -1 \right) \rho \Omega(\rho,T),
\end{equation}
\begin{equation}\label{state}
    p = \frac{k_B}{\mu m_H} \rho T,
\end{equation}
where the mass density $\rho$, the enclosed mass $M$, the radial
flow velocity $u$, the thermal gas pressure $p$, and the temperature
$T$, in general, depend on the radius $r$ and time $t$; $G$ is the
gravitational constant, $\gamma$ is the polytropic index, and $k_B$,
$\mu \approx 2.3$ and $m_H$ are Boltzmann constant, the mean
molecular weight and the hydrogen mass, respectively. The net
cooling function is represented by $\Omega(\rho,T)$.

In the stationary ($\partial/\partial t =0$) quasi-static
($u\rightarrow 0$) thermally equilibrium state, the net cooling
function $\Omega(\rho,T)$ must be zero at each radius $r$ (i.e.,
locally thermal balance). To calculate the thermal balance within
the molecular cloud cores, we need to consider heating and cooling
processes affecting the gas and the dust. For cooling processes, we
use the Table~2 of Goldsmith~(2001) to parameterize the cooling
function as $\Lambda_0 \left( T/10\,\mathrm{K} \right)^\beta$ where
$\Lambda_0$ and $\beta$ are generally functions of density (see,
Fig.~1 of Nejad-Asghar~2011). There are several different heating
mechanisms in models of interstellar matter. Here, we consider
heating processes of cosmic rays (e.g., Glassgold \& Langer~1973)
and dissipation of magnetic energy via AD (e.g., Li, Myers \&
McKee~2012).

The heating due to cosmic rays with sufficient energies ($\sim 100
\mathrm{MeV}$) to penetrate dense clouds is commonly about
$\Gamma_{CR} = 2.5 \times 10^{-4} \mathrm{erg\,g^{-1}\,s^{-1}}$,
with assumption of an ionization rate per $H_2$ molecule of $2\times
10^{-17} \mathrm{s^{-1}}$ and a mean energy gains per ionization of
$19 \mathrm{eV}$ (Glassgold \& Langer~1973). The heating due to AD
can be obtained by low ionization fraction approximation. In the
limit of low ionization fraction (i.e., $\rho = \rho_n + \rho_i
\approx \rho_n$, where $\rho_n$ and $\rho_i$ are neutral and ion
densities, respectively), the inertia of charged particles being
negligible so that the Lorentz force will be balanced by the equally
important drag force. The drag force per unit volume exerted on the
neutrals by ions is $\textbf{f}_d = \gamma_{AD} \rho_i \rho
\textbf{v}_d = \gamma_{AD} \epsilon \rho^{3/2} \textbf{v}_d$, where
$\gamma_{AD} \sim 3.5 \times 10^{13} \mathrm{cm^3\,g^{-1}\,s^{-1}}$
is the collisional drag coefficient in the molecular clouds, and we
used the relation $\rho_i=\epsilon\rho^{1/2}$ between ion and
neutral densities in the local ionization equilibrium state with
$\epsilon\sim 3\times 10^{-16} \mathrm{g^{1/2}\,cm^{-3/2}}$
(Shu~1992). In this way, the drift velocity of ions is
\begin{equation}\label{drift}
    \textbf{v}_d = \textbf{v}_i - \textbf{v}_n \approx \frac{1}{4\pi \gamma_{AD} \epsilon\rho^{3/2}}(\nabla\times\textbf{B})\times
    \textbf{B}.
\end{equation}
The magnitude of drift velocity, $v_d$, is inversely proportional to
the power of density and directly proportional to the magnetic field
strength and its gradient. Shu~(1992, Equation~27.9) used typical
values of $B \Delta B/\Delta x \sim (30 \mu \mathrm{G})^2 / 0.1
\mathrm{pc}$ to estimate the typical drift speeds through the
molecular clouds. Li, Myers \& McKee~(2012) divided the magnetic
field into a steady component and a fluctuating one, and used the
mean squared method to estimate upper limits on the drift speed.
Here, we use the parametric relation $v_d = B \kappa \rho^{-3/2}/
4\pi \gamma_{AD} \epsilon$, where the parameter $\kappa \equiv
\Delta B/ \Delta x$ is the change of magnetic field strength in the
length-scale $\Delta x$. In this way, the heating due to ambipolar
diffusion can be represented as
\begin{equation}\label{heatingAD}
  \Gamma_{AD} = \frac{\textbf{f}_d.\textbf{v}_d}{\rho} =
  \frac{B^2\kappa^2}{16\pi^2\gamma_{AD} \epsilon}
  \rho^{-\frac{5}{2}}.
\end{equation}

The magnetic field strength, $B$, is evaluated in the Troland \&
Crutcher~(2008) for a set of $34$ molecular cloud cores. Their
evaluations show that the magnetic field strengths are in the range
of $0.5$ to $50\,\mathrm{\mu G}$. The authors use a scaling relation
of the field strength with density, which is usually parameterized
as a power law, $B\propto \rho^\eta$ (Crutcher~2012). In the
strong-field models, $\eta \lesssim 0.5$ is predicted (e.g.,
Mouschovias \& Ciolek~1999), while for weak magnetic fields, $\eta
\approx 0.66$ is predicted (Mestel~1966). Here, we assume that the
magnetic field is so strong that as ambipolar diffusion increases
the mass-to-flux ratio, the density $\rho$ increases faster than
$B$. Thus, we are in strong-field models with $\eta \lesssim 0.5$.
According to the Figure~6 of Crutcher~(2012), we choose an
approximate value of $B\approx12\,\mathrm{\mu G}$ for number density
$n=\rho/\mu m_H \approx 10^3\, \mathrm{cm}^{-3}$, and
$B\approx100\,\mathrm{\mu G}$ for number density $n=\rho/\mu m_H
\approx 10^6\, \mathrm{cm}^{-3}$. By choosing these values for the
magnetic field strengths, we have a power-law approximation as
$B\approx 100\, \mathrm{\mu G} \left(\frac{n}{10^6 \,
\mathrm{cm}^{-3}}\right)^{0.3}$. In this way, the condition for
domination of thermal pressure over the magnetic pressure is
$\log(n/\mathrm{cm}^{-3})\gtrsim 7.25 - 2.5\log(T/\mathrm{K})$.
According to the Table~1 of Koch, Tang \& Ho~(2010), the maximum
value of $\Delta B$ is $\sim 0.9 B$, which corresponds to cloud core
with dimension $60\,\mathrm{mpc}$ at W51~e2/e8, observed with SMA
interferometer. On the other hand, like the work of Li, Myers \&
McKee~(2012), we assume that the AD heating, which is produced by
magnetic fluctuations, is wave dissipation with its maximum
wavelength equal to the physical scale $l_0$. In this way, gradients
cannot exist on scales larger than $l_0/2\pi$. Here, we choose the
minimum value of $l_0$ equal to $23\,\mathrm{mpc}$, which
corresponds to the cloud core at Orion~BN/KL, observed with SMA
interferometer (Koch, Tang \& Ho~2010, Table~1), so that we have
$\Delta x \sim 3\,\mathrm{mpc}$. Thus, we expect the maximum value
of the parameter $\kappa$ in the prestellar cores be equal to $\sim
3\, \mathrm{\mu G} / 1\, \mathrm{mpc}$. Since there is a lot of
uncertainties in the values of the magnetic field fluctuations and
dissipation scales, we assume a parametric value for $\kappa$, which
includes roughly all of these uncertainties. We express $\kappa$ in
the unit of $0.3 \, \mathrm{\mu G} / 1\, \mathrm{mpc}$ and perform
the calculations for $0\lesssim \kappa \lesssim 10 \times 0.3\,
\mathrm{\mu G} / 1\, \mathrm{mpc}$. Thus, the AD heating
(\ref{heatingAD}) can be rewritten as
\begin{equation}\label{adheating}
    \Gamma_{AD} = 2.3 \times 10^{-9}
    \left( \frac{\kappa}{0.3\, \mathrm{\mu G}/1 \mathrm{mpc}} \right)^2
    \left( \frac{n}{10^{6} \mathrm{cm}^{-3}}
     \right)^{-1.9}\, \mathrm{erg\,g^{-1}\,s^{-1}}.
\end{equation}

\begin{figure} \epsscale{0.8} \plotone{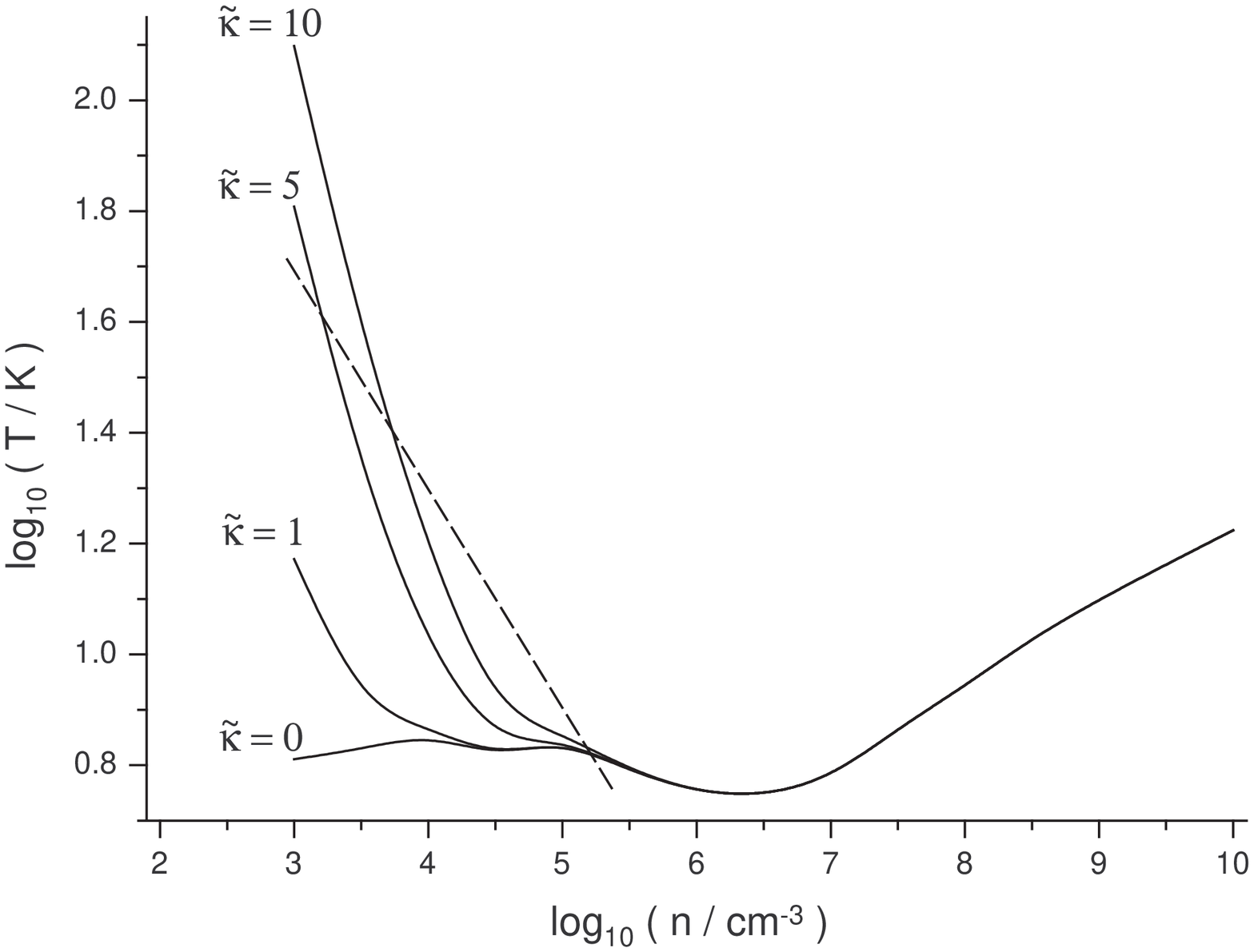}
\caption{The temperature-density relation for thermal balancing
state of the molecular clouds. The parameter $\tilde{\kappa}\equiv
\Delta B/ \Delta x$ represents the change of magnetic field strength
at length-scale $\Delta x$, in unit of $0.3\,\mathrm{\mu G}/1
\mathrm{mpc}$. The dash-line is $\log(n/\mathrm{cm}^{-3})= 7.25 -
2.5\log(T/\mathrm{K})$ in which for densities grater than this
value, the assumption of thermally dominated prestellar core is
approximately correct.\label{tempfig}}
\end{figure}

The thermal balance (i.e., $\Omega(\rho,T) = \Lambda_0(n) \left(
T/10\,\mathrm{K} \right)^{\beta(n)} -\Gamma_{CR} - \Gamma_{AD} = 0$)
at the molecular clouds leads to a relation between the temperature
and density as follows
\begin{equation}\label{temp}
    T = 10\,\mathrm{K} \times \left[ \frac{\Gamma_{CR}}{\Lambda_0(n)} +
    \frac{2.3\times10^{-9}}{\Lambda_0(n)}
    \left( \frac{\kappa}{0.3\,\mathrm{\mu G}/1\, \mathrm{mpc}} \right)^2
    \left( \frac{n}{10^{6} \mathrm{cm}^{-3}}
     \right)^{-1.9} \right]^{\frac{1}{\beta(n)}}.
\end{equation}
The results of (\ref{temp}) are plotted in the Fig.~\ref{tempfig},
for different values of $\tilde{\kappa} \equiv \kappa / (\frac{0.3\,
\mathrm{\mu G}}{1\,\mathrm{mpc}})$. The dash-line in this figure,
$\log(n/\mathrm{cm}^{-3})= 7.25 - 2.5\log(T/\mathrm{K})$, shows that
our assumption of thermally dominated prestellar core is
approximately correct, especially for greater values of $\kappa$.
Departure from the thermally dominated assumption leads to
consideration of the pressures from both MHD turbulence and the mean
field gradient; the resulting equilibrium configuration will not be
spherical. Here, we use the spherical assumption for prestellar
cores as a first approximation, and the effects of magnetic
pressures will be considered in the subsequent work.

Now, knowing the relation between temperature and density, equation
(\ref{state}) leads us to determine the gradient of pressure as
\begin{equation}\label{dpdr}
    \frac{dp}{dr} = \frac{k_B}{\mu m_H} \left( T+\rho\frac{dT}{d\rho} \right)
    \frac{d\rho}{dr},
\end{equation}
so that the stationary ($\partial/\partial t =0$) quasi-static
($u\rightarrow 0$) state of the momentum equation (\ref{moment})
becomes
\begin{equation}\label{hydrost}
    \frac{d\rho}{dr} = - \frac{\mu m_H G}{k_B}
    \frac{M\rho}{r^2 \left( T+ \rho\frac{dT}{d\rho} \right)}.
\end{equation}
We use the non-dimensional quantities $\tilde{n} \equiv n/n_r$,
$\tilde{T} \equiv T/T_r$, $\tilde{r} \equiv r/(\frac{k_BT_r/\mu
m_H}{4\pi G \mu m_H n_r})^{\frac{1}{2}}$, and $\tilde{M} \equiv M/
4\pi (\frac{k_BT_r/\mu m_H}{4\pi G \mu m_H n_r})^{\frac{3}{2}} \mu
m_H n_r$, where $n_r=10^6\, \mathrm{cm}^{-3}$ and
$T_r=10\,\mathrm{K}$ are the reference density and temperature,
respectively. In this way, the equations (\ref{mass}) and
(\ref{hydrost}) become
\begin{equation}\label{hydmass}
    \frac{d\tilde{M}}{d \tilde{r}} = \tilde{r}^2 \tilde{n},
\end{equation}
\begin{equation}\label{hydmoment}
    \frac{d\tilde{n}}{d\tilde{r}} = - \frac{\tilde{M}\tilde{n}}
    {\tilde{r}^2 \left( \tilde{T}+ \tilde{n}\frac{d\tilde{T}}{d\tilde{n}}
    \right)},
\end{equation}
which can be integrated numerically (e.g., with Runge-Kutta method),
from the origin $\tilde{r}=0$ with the boundary conditions
$\tilde{n}(0)=\tilde{n}_c$ and $\tilde{M}(0)=0$, where $\tilde{n}_c$
is the non-dimensional central density.

\begin{figure} \epsscale{0.8} \plotone{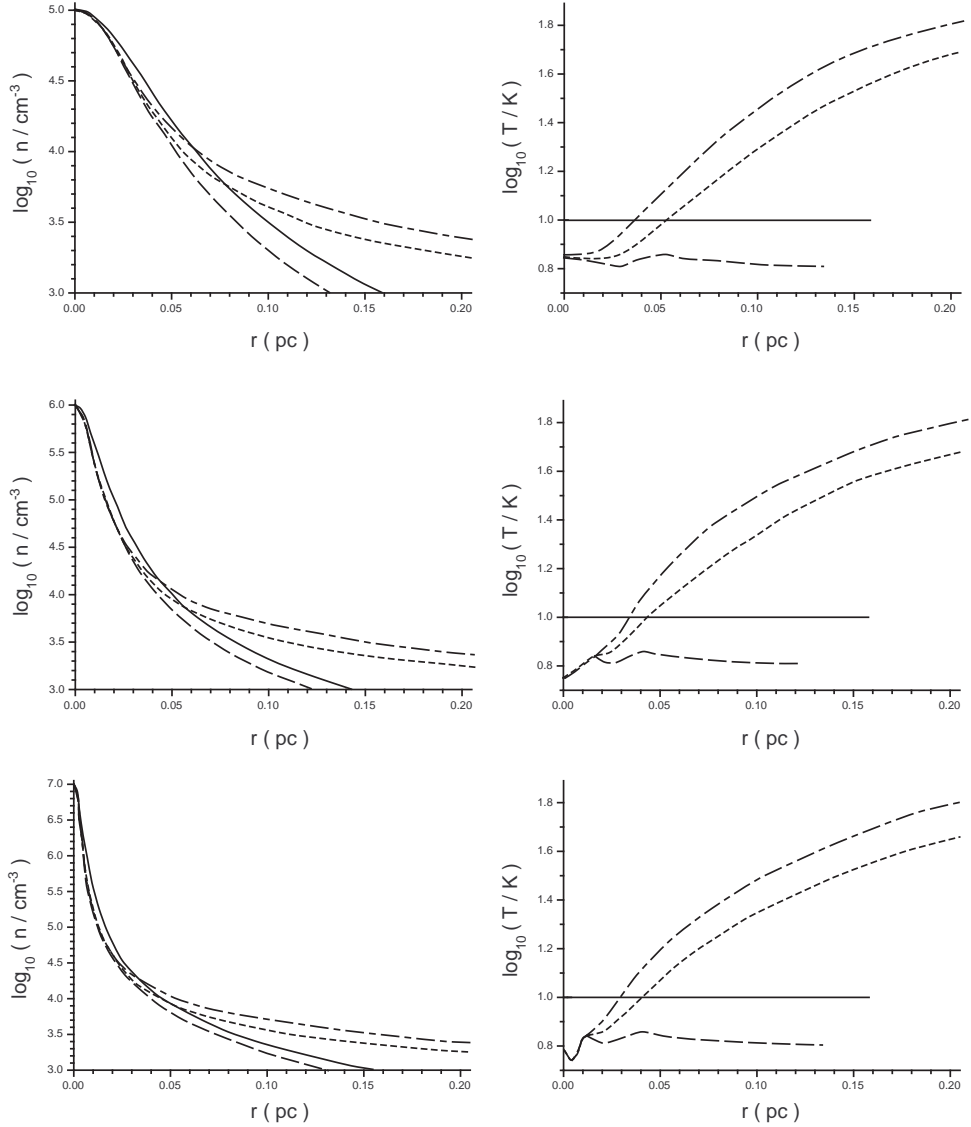}
\caption{The density and temperature profiles of dynamically
balanced, spherical molecular cloud cores in the isothermal case
(solid), and in the non-isothermal cases with $\tilde{\kappa}=0$
(dash), $5$ (dot), and $10$ (dash-dot), for three values of central
density equal to $10^5\,\mathrm{cm}^{-3}$ (upper-panels),
$10^6\,\mathrm{cm}^{-3}$ (middle-panels), and
$10^7\,\mathrm{cm}^{-3}$ (lower-panels).\label{dentempradius}}
\end{figure}

The results for density and temperature profiles of the stationary
quasi-static thermally equilibrium prestellar cores with central
density $\tilde{n}_c=0.1$, $1$, and $10$ are shown in the
Fig.~\ref{dentempradius}. The profiles of isothermal case
($\tilde{T}=\mathrm{const.}$) are also plotted in this figure, for
comparison. As can be seen, consideration of the AD heating departs
the profiles of the stationary quasi-static thermally equilibrium
prestellar cores from the isothermal case. These departures are
greater for greater magnetic fluctuating parameter $\kappa$, at the
envelopes of the prestellar cores with smaller densities. This
result for departure of profiles is consistent with the
Fig.~\ref{tempfig}, which shows that importance of AD heating is at
lower densities.

\begin{figure} \epsscale{0.5} \plotone{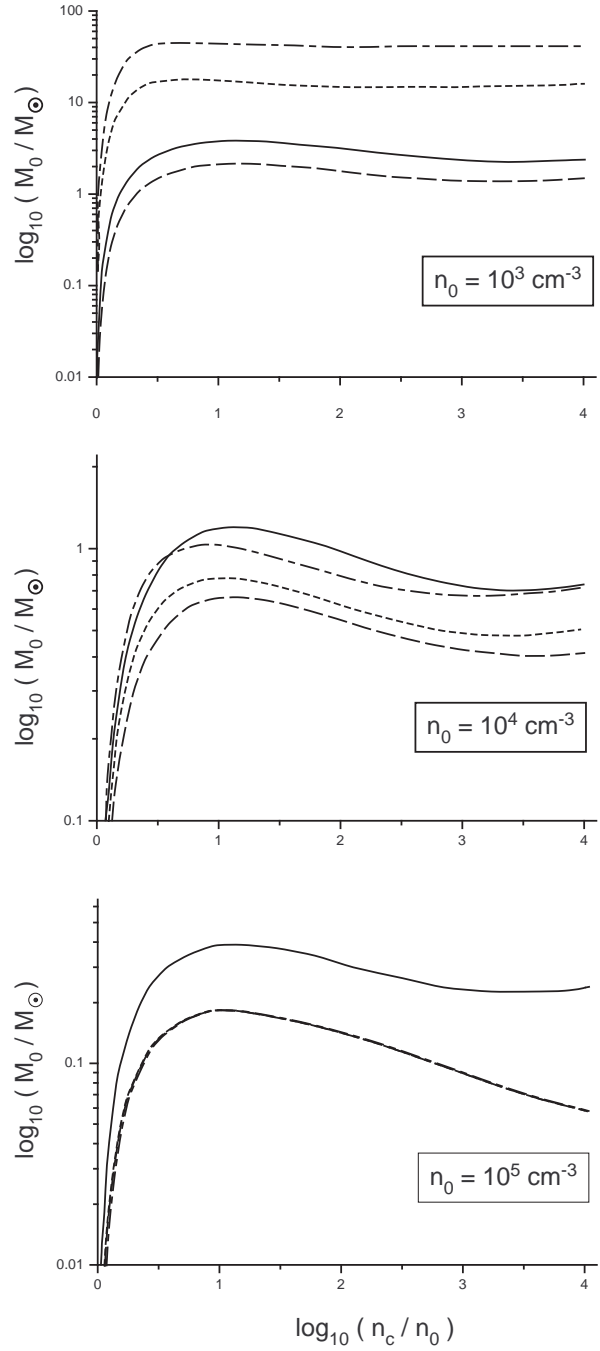}
\caption{The core mass versus density contrast for fixed external
densities $n_0=10^3\,\mathrm{cm}^{-3}$ (upper-panel),
$10^4\,\mathrm{cm}^{-3}$ (middle-panel), and
$10^5\,\mathrm{cm}^{-3}$ (lower-panel), for isothermal case (solid),
and non-isothermal cases with $\tilde{\kappa}=0$ (dash), $5$ (dot),
and $10$ (dash-dot). In the lower-panel, the non-isothermal cases
with different $\tilde{\kappa}$ overlap to each
other.\label{massfig}}
\end{figure}

In the actual cases, the density does not fall to zero, but to some
value $\tilde{n}_0$ characterizing the external medium. If we fix
$\tilde{n}_0$, for each density contrast $\tilde{n}_c /
\tilde{n}_0$, we can find the radius $\tilde{r}_0$ of the prestellar
core and its corresponding non-dimensional mass $\tilde{M}_0$. The
molecular cloud core mass, $M_0$, versus logarithm of density
contrast, $n_c/n_0$, are plotted in Fig.~\ref{massfig}, for three
different values of density of external medium equal to $n_0=10^3$,
$10^4$, and $10^5 \mathrm{cm}^{-3}$. With increasing density
contrast, core mass first rises to a maximum value of $M_{MBES}$,
attained at the radius $r_{MBES}$. Then, the mass drops to a
minimum, and eventually approaches in an oscillatory fashion to the
asymptotic limit value, which represents the mass of the singular
sphere (with density $\propto 1/r^2$). In the isothermal
Bonnor-Ebert case, the density contrast corresponded to the first
maximum is $n_c / n_0=14.1$ (e.g., Stahler \& Palla~2004,
Figure~9.2).

\begin{figure} \epsscale{0.6} \plotone{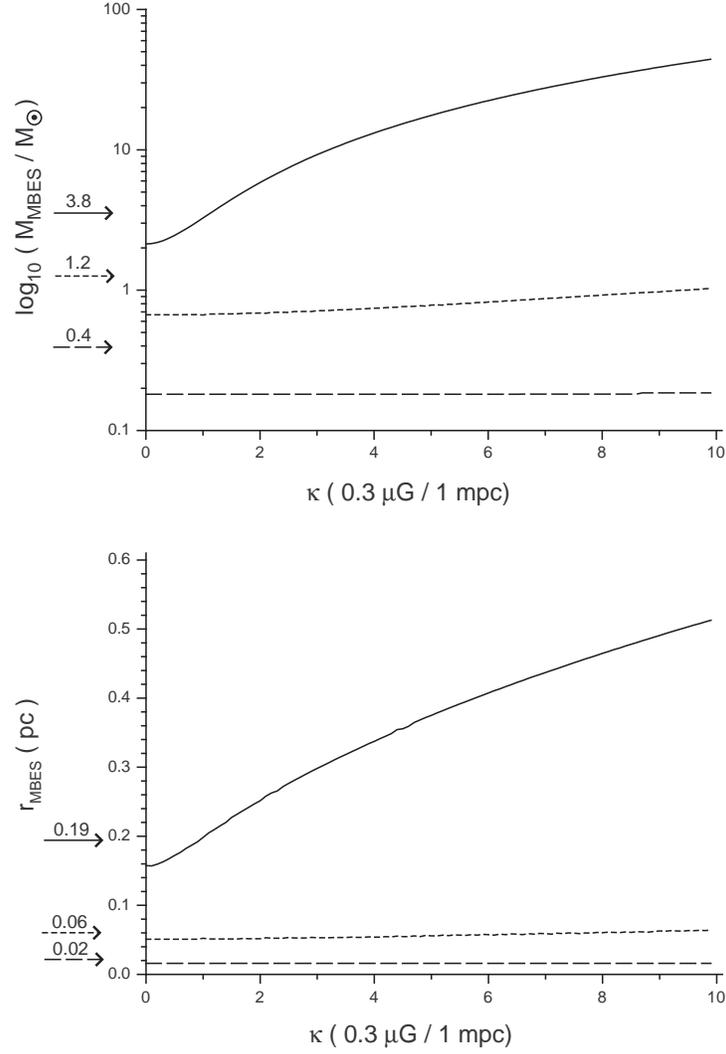}
\caption{The mass and radius of MBES versus magnetic fluctuation
parameter $\kappa$, for fixed external density
$n_0=10^3\,\mathrm{cm}^{-3}$ (solid), $10^4\,\mathrm{cm}^{-3}$
(dot), and $10^5\,\mathrm{cm}^{-3}$ (dash). The corresponded
isothermal results are depicted by arrows on the vertical
axes.\label{rmbefig}}
\end{figure}

Prestellar cores with low density contrast are mainly confined by
the external pressure, and not self-gravity. This situation changes
as we progress along the curves in the Fig.~\ref{massfig} to the
right of the first maximum (i.e., models of higher $n_c/n_0$). These
prestellar cores have gravity-dominated configurations, and an
arbitrarily small initial perturbation in the structure grows
rapidly with time, leading ultimately to collapse. In the isothermal
case, a cloud with the critical value of mass and radius
corresponding to the first maximum is known as the Bonnor-Ebert
sphere. For non-isothermal cases, we use the nomenclature of
modified Bonnor-Ebert sphere (MBES). The critical values of mass and
radius of the MBESs, versus $\kappa$, are plotted in the
Fig.~\ref{rmbefig}, for different values of external densities.

\section{Summary and conclusions}

The fluctuations of the magnetic field strength through the
molecular clouds can produce AD heating. In this research, the
effects of AD heating to produce temperature gradient through
thermally supported, dynamically balanced prestellar spherical cores
are investigated. This temperature gradient changes the maximum
allowed mass and radius of the gravitationally stable cores (i.e.,
MBES). The issue of MBES is important in the star formation theory.

Thermal balance of the molecular clouds demand forcefully a relation
between density and temperature as outlined in Fig.~\ref{tempfig},
for different values of magnetic fluctuation parameter $\kappa$.
From thermal balancing state (Fig.~\ref{tempfig}), we know the
relation between temperature and density, so that we can find both
temperature $T(r)$ and density $\rho(r)$ in which the stationary
hydrostatic equilibrium of prestellar cores are satisfied. The
results for temperature and density profiles are shown in the
Fig.~\ref{dentempradius}. The inclusion of AD heating leads to the
departure of the density and temperature profiles from the
isothermal case, especially in the outer regions of envelope with
lower densities. This effect leads to the formation of the
gravitationally stable high-mass prestellar cores in the low density
medium.

The Fig.~\ref{massfig} depicts the core mass as function of density
contrast. For fixed external density $n_0$, increasing of central
density leads to rise of core mass to a maximum value of $M_{MBES}$
attained at the critical radius $r_{MBES}$. The prestellar cores
with masses lower than $M_{MBES}$ are gravitationally stable, while
if the core mass reaches to the critical value $M_{MBES}$ or a
larger value, the core will be gravitationally unstable and collapse
occurs. The Fig.~\ref{massfig} shows that in the high-density
medium, ambipolar diffusion is not important, while in the
low-density medium, it is important so that we can have high-mass
cores that are gravitationally stable.

Importance of magnetic fluctuation parameter $\kappa$ to produce
high-mass prestellar cores at low-density molecular clouds are shown
in Fig.~\ref{rmbefig}. In isothermal case, the critical density
contrast for first maximum is attained at $n_c / n_0 = 14.1$, while
in the non-isothermal cases without AD heating ($\kappa=0$), this
value increases so that its corresponding critical radius $r_{MBES}$
lies below the isothermal value, as outlined by Sipil\"{a}, Harju \&
Juvela~(2015). Considering of magnetic fluctuation and AD heating
($\kappa\neq 0$) increases the $r_{MBES}$, and leads to existence of
gravitationally stable high-mass prestellar cores at the low-density
molecular medium.

\section*{Acknowledgments}
I appreciate the the anonymous reviewer for his/her careful reading,
useful comments and suggested improvements.




\clearpage

\clearpage

\clearpage

\clearpage

\end{document}